\documentclass{aa}
\usepackage{graphicx,amsfonts}
\begin{document}
   \title{Dust formation in winds of long-period variables}

   \subtitle{V. The influence of micro-physical dust properties in carbon stars}

   \author{Anja C.\,Andersen\inst{1,}\inst{2}
          \and
          Susanne H\"ofner\inst{1}
          \and
          Rita Gautschy-Loidl\inst{3}
          }

   \offprints{A.C.\,Andersen}

   \institute{Department of Astronomy \& Space Physics, 
     Uppsala University, P.O.Box 515, SE-751 20 Uppsala, Sweden\\
   \email{hoefner@astro.uu.se}
   \and 
     Astronomical Obs., NBIfAFG, Copenhagen University,
     Juliane Maries Vej 30, DK-2100 Copenhagen, Denmark\\
   \email{anja@astro.ku.dk} 
   \and 
     R\"uti, Switzerland \\
     \email{rita@gautschy.ch}
   }
   \date{Received 21-10-02; accepted 07-01-03}

   \abstract{We present self-consistent dynamical models for dust-driven winds
of carbon-rich AGB stars. The models are based on the coupled system of frequency-dependent 
radiation hydrodynamics and time-dependent dust formation. We investigate in detail
how the wind properties of the models are influenced by the micro-physical properties of the
dust grains that are required by the description of grain formation.  The choice of dust parameters is 
significant for the derived outflow velocities, the degrees of condensation and the 
resulting mass-loss rates of the models.
In the transition region between models with and without mass-loss the choice of
micro-physical parameters turns out to be very significant for whether a particular set of stellar
parameters will give rise to a dust-driven mass-loss or not. We also calculate near-infrared colors to test
how the dust parameters influence the observable properties of the models, however, at this point we do not attempt to fit particular stars.
    }
   \authorrunning{Andersen et al.} 
   \titlerunning{Dust formation in winds of long-period variables}
   \maketitle
%

\section{Introduction}

Mass loss by dust-driven winds of asymptotic giant branch (AGB) stars is
probably one of the major mechanism which recycle 
material in the Galaxy (e.g.\ Sedlmayr \cite{sedlmayr94}). Most stars 
($M_{\star} \le 8 M_{\odot}$) will eventually become AGB stars 
and subsequently end their life as white dwarfs surrounded by planetary nebulae. 

AGB stars are cool ($T_{\star} < 3500$~K) and luminous ($L_{\star}$ of a few
$10^{3}$ to a few $10^{4}$~L$_{\odot}$),
and a majority of them are pulsating long-period variables (LPVs). 
The outer layers of many AGB stars provide favorable conditions for 
the formation of molecules and dust grains. Dust grains play an
important role for the heavy mass-loss (up to 
$\dot{M} \sim 10^{-4} M_{\odot}$/yr) of these stars by transferring
momentum from the radiation to the gas. 
Pulsation causes an extended atmosphere where the dust is
condensing. The dust absorbs the 
light of the central star and re-radiates it at longer wavelengths in the
infrared range ($\lambda > 2 \mu$m). The density of material in the
circumstellar envelope may be so large that the star completely disappears
in the visual range. 

Considerable effort has been put into
a theoretical description of the mass-loss of AGB stars. 
At the end stages of stellar evolution the mass-loss rates become
so high that they (and not the nuclear burning rates) determine
the stellar evolutionary faith.
The goal is to develop a mass-loss description that can be used as input to
models of stellar evolution and the chemical evolution of the Galaxy.

This paper is the fifth in our series on dust formation in winds of 
long-period variables.  
The previous models presented in Paper I-IV  
(Dorfi \& H\"ofner \cite{dorfi+hoefner91}; H\"ofner \& Dorfi \cite{hoefner+dorfi92};
H\"ofner et al.\ \cite{hoefner+etal95}; H\"ofner \& Dorfi \cite{hoefner+dorfi97})
are all based
on gray radiative transfer while the models presented here are
calculated using frequency dependent radiative transfer for the
gas and dust (see H\"ofner et al.\ \cite{hoefner+etal02} for details).  
In this paper we focus on the influence of the micro-physical 
(i.e.\ optical and chemical) properties of dust grains on the winds of AGB 
stars, to determine to what extent the choice of micro-physical parameters 
effects the general mass-loss predictions. 

In Sect.\,2 we discuss the hydrodynamical
models with emphasis on the treatment of the grain formation. 
Section 3 describes the different types of amorphous carbon dust used 
to test the model dependence on the choice of the micro-physical parameters.
In Sect.\,4 we present the results and show that the choice of opacity data,
the values used for the sticking coefficients and the intrinsic dust density 
of the material, all affect the mass-loss rates and other wind properties
resulting from the models. This has an influence on the calculated 
synthetic colors which are compared with observations. 
Conclusions are presented in Sect.\,5.


\section{The hydrodynamical model} \label{model}

Our spherically symmetric hydrodynamical models predict the mass-loss rates of
AGB stars by treating in detail the atmosphere and the circumstellar 
environment around pulsating long-period variable stars.
This is done by solving the coupled system of frequency-dependent radiation hydrodynamics and
time-dependent dust formation (cf. H\"ofner et al.\ \cite{hoefner+etal02})
employing an implicit numerical method and an adaptive grid
(for details on the numerical technique see Dorfi \& Feuchtinger \cite{dorfi+feuchtinger95}).


\subsection{Hydrodynamics and radiative transfer}

In our radiation-hydrodynamical models the stellar and circumstellar 
envelope are described in terms of separate 
conservation laws for the dust, the gas and the radiation field. 
The resulting set of nonlinear partial differential equations consist of: 
\begin{itemize}
\item Equation of continuity (mass conservation).
\item Equation of motion (matter momentum conservation).
\item Equation of gas internal energy (matter energy conservation).
\item 0. moment of the radiative transfer equation (radiation energy conservation).
\item 1. moment of the radiative transfer equation (radiation momentum conservation).
\item Moment equations for the dust (grain formation and grain growth).
\item Poisson equation (self gravitation).   
\end{itemize}
To this system of nonlinear partial 
differential equations we add the so-called grid equation which determines
the locations of the grid points, depending on some critical physical
quantities to be resolved during the computations. 

For the dust component only certain moments of the grain size 
distribution have to be known for a complete description of the
circumstellar envelope (more details in Sect.\,\ref{dustmoment}).

In contrast to earlier models in this series 
the models presented here describe the radiation field by a
frequency-dependent treatment of the gas and
dust (H\"ofner \cite{hoefner99}; H\"ofner et al.\ \cite{hoefner+etal02})
based on opacity sampling data of
molecular opacities (SCAN data base, J{\o}rgensen \cite{joergensen97})
at 51 frequency points between 0.25 and 12.5 $\mu$m. Solving the
frequency-dependent transfer increases the computation time per
time step considerably.
To keep the computation time at a reasonable level,
the spatial grid points available in the model are reduced from 
500 to 100 compared to the gray models presented in e.g.\ H\"ofner \& Dorfi 
(\cite{hoefner+dorfi97}).

Basically four 
stellar parameters are needed as input; the stellar mass ($M_{\star}$),
the effective temperature ($T_{\rm eff}$), the luminosity
($L_{\star}$) of the star and the gas abundance ratio 
(${\varepsilon_{\rm C}/\varepsilon_{\rm O}}$) of carbon
to oxygen.  However, to describe the pulsation of the star, 
two additional parameters are required. 
Pulsation is simulated by a sinusoidal motion
of the inner boundary $R_{in}$ which is located below the stellar
photosphere.  This variable
inner boundary is parameterized by a velocity amplitude ($\Delta u$)  
and a pulsation period ($P$). The luminosity at the inner boundary is 
variable and no mass flux is allowed across this boundary
(cf.\ H\"ofner \& Dorfi \cite{hoefner+dorfi97}). 
As in previous models a perfect gas law with $\gamma = 5/3$ and 
$\mu = 1.26$ is used as equation of state for reasons of comparability. 

The dynamical calculations start from a dust free hydrostatic initial model.
The radiative pressure on newly formed dust initiates an outward motion and the
expansion is followed by the grid out to around 20--30~$R_{\star}$. 
At this radius the outer boundary is fixed, allowing for
outflow. A model evolves for typically 100 years.
The model calculation is stopped before a significant depletion of 
the mass inside the computational domain occurs.


\subsection{Grain formation in the models}

Dust influences the dynamics and thermodynamics of the stellar
atmosphere by its opacity. To calculate this opacity we need
to know the amount of dust present. Dust formation proceeds far from
equilibrium and it is necessary to use a 
detailed time-dependent description to determine the degree of 
condensation and other relevant properties of the grains.

Dust is formed by a series of chemical reactions in which atoms or
molecules from the gas phase combine to clusters of increasing
size. The molecular composition of the gas phase
determines which atoms and molecules are available for the cluster
formation and grain growth. 
Dust formation begins with nucleation of critical clusters 
followed by growth to macroscopic dust grains.  


\subsubsection{Grain growth} \label{dustmoment}

In our models grain formation is treated by the so-called moment method
(Gail \& Sedlmayr \cite{gail+sedlmayr88}; Gauger et al.\ \cite{gauger+etal90}).
The moment method describes the
time evolution of an ensemble of dust grains of various sizes (including
effects due to chemical and thermodynamical non-equilibrium) and
requires the nucleation rate as external input. 

In the moment method the size of
a single dust grain is expressed in terms of the number of monomers ($N$) 
contained in the grain. A monomer represents the basic element the 
grain is built of, i.e. a
certain species of atoms or molecules. The ensemble of the grains is described by the
size distribution function $f(N,t)$ representing the number densities of dust particles
in dependence of their size $N$. The number density $f(N,t)$ of dust grains containing
$N$ monomers is changed by four processes: creation of grains of size $N$ by growth of smaller
dust particles and by destruction of larger ones as well as destruction of grains with
$N$ monomers by growth or evaporation (see Gail \& Sedlmayr \cite{gail+sedlmayr88} and
Gauger et al.\ \cite{gauger+etal90} for details).

As long as the size of the dust grains is small compared to the 
wavelength of the photons in the relevant frequency region, 
which is the case for late-type stars, the optical properties
of the dust do not depend explicitly on the size spectrum $f(N,t)$
but only on a few moments ($K_{j}$) of the grain size distribution 
function defined by
\begin{equation}
K_{j} = \sum_{N = N_{l}}^{\infty} N^{j/d} \, f(N,t),
\end{equation}
where $d$ denotes the spatial dimension of the grains ($d=3$ for spherical particles and 2 for planar structures) and
$N_{l}$ is the lower limit of the grain sizes which may be regarded as macroscopic
in the thermodynamical sense.  In the work presented here we
use the value $d=3$ (i.e.\ assume spherical grains) and $N_{l} = 1000$.

From the moments $K_{j}$ it is possible to calculate quantities like the total
number density of dust grains, the mean grain radius and the fraction of condensible material 
actually condensed into grains.

Considering grains large enough that their thermodynamical properties do not depend
on the grain size and assuming that only molecules with up to a few monomers 
contribute significantly to the growth process the following set of moment
equations can be derived (Gauger et al.\ \cite{gauger+etal90})
\begin{eqnarray}
\frac{dK_{0}}{dt} & = & {\cal J} \label{growth} \\
\frac{dK_{j}}{dt} & = & \frac{j}{d} \, \frac{1}{\tau} \, K_{j-1} + N_{l}^{j/d} \, {\cal J}  
                        \qquad (1 \le j \le d). \label{dk/dt}
\label{growth2}
\end{eqnarray}
Here Eq.\,(\ref{growth}) determines the grain production rate.
The quantity $1/\tau$ is the net growth rate of the dust grains and $\cal J$ is the net transition rate
per volume from cluster sizes $N<N_{l}$ to $N>N_{l}$, this can
be interpreted as the local current density of clusters flowing up-wards in cluster 
size space from the region $N \le N_{l}$ (Gail \& Sedlmayr \cite{gail+sedlmayr88}).

The net growth rate $1/\tau$ contains the number densities of the chemical
species which take part in the dust formation process. 
In our models, the relevant quantities
are obtained by assuming chemical equilibrium in the gas phase at the gas
temperature $T_{\rm g}$
\begin{eqnarray}
\left( \frac{1}{\tau} \right)_{\rm CE} & =& 
\sum_{i = 1}^{I} i \, A_{1} \, v_{th} (i) \,
   \alpha (i) \, f(i,t) \, \\ \nonumber 
& &  \left[  1 - \frac{1}{{\cal S}^{i}} \frac{{\cal K}_{i} (T_{\rm d})}{{\cal K}_{i} (T_{\rm g})} \, \sqrt{\frac{T_{\rm g}}{T_{\rm d}}} \, \right] \\ \nonumber 
   &+& \sum_{i = 1}^{I'} i \, A_{1} \sum_{m = 1}^{M_{i}} v_{th} (i,m) \, \alpha_{m}^{c} (i)\,  n_{i,m} \, \\ \nonumber & &   
\left[  1 - \frac{1}{{\cal S}^{i}} \frac{{\cal K}_{i,m}^{r}(T_{\rm g})}{{\cal K}_{i,m}^{r}(T_{\rm d})} \, \frac{{\cal K}_{i,m}(T_{\rm d})}{{\cal K}_{i,m}(T_{\rm g})} \, \right] 
\label{tau}
\end{eqnarray}
where $\cal K$ denotes the dissociation constant of the molecule of the growth reaction and ${\cal K}^{r}$ is the 
dissociation constant of the molecule involved in the reverse reaction (see Gauger et al.\ \cite{gauger+etal90}).
$f(i)$ and $n_{i,m}$ are the number densities of the $i$-mers and the molecules containing $i$-mers which 
contribute to the grain growth, respectively. $A_{1}$ denotes the (hypothetical) monomer surface area $A_{1}=4 \pi a_{1}^2$ 
where $a_{1}$ is the monomer radius (see Eq.\,(\ref{monomerradius}), $v_{th}$ the thermal
velocities of the corresponding growth species and $\alpha$ the sticking coefficients. 
The supersaturation ratio $\cal S$ is defined by
\begin{equation}
{\cal S} = \frac{P_{\rm mon}}{P_{\rm sat}(T_{\rm d})},
\label{supersaturation}
\end{equation}
which is the ratio of the actual partial pressure of the monomers in the gas phase to the vapor
saturation pressure with respect to the dust temperature $T_{\rm d}$. 
If the thermodynamical conditions allow the formation of dust grains the net
transition rate ${\cal J}$ is assumed to be equal to the nucleation rate, i.e.\ the rate at which supercritical
(stable) clusters are formed out of the gas phase (see Sect.\,\ref{nucleate} for details).

The number densities of the molecules relevant 
to the dust formation are calculated assuming chemical equilibrium
between H, H$_{2}$, C, C$_{2}$, C$_{2}$H and C$_{2}$H$_{2}$ after
the fraction of carbon bound in CO has been subtracted. Nucleation,
growth and destruction of dust grains are supposed to proceed
by reactions involving C, C$_{2}$, C$_{2}$H and C$_{2}$H$_{2}$.
The values of the elemental abundances are taken to be solar 
(Allen \cite{allen73}) except for the carbon abundance which is considered
as a free parameter. The dissociation constants ${\cal K}(T)$ have
been extracted from the JANAF tables (Stull \& Prophet \cite{stull+prophet71}).


\subsubsection{Grain nucleation} \label{nucleate}

Nucleation is the first stage of the condensation process whereby a vapor transforms
to a solid or liquid. This phase change requires some degree of supersaturation 
in order to drive the system through the relatively unstable reactive intermediates (clusters)
between the atomic or molecular vapor and the macroscopic solid or liquid states.
Presently no nucleation rates based on calculations of chemical pathways are available 
for the astrophysical problem under consideration here. Consequently, the nucleation rate which
is a function of temperature, density and supersaturation ($\cal S$) for a particular vapor is often
calculated by either the classical homogeneous nucleation theory (Becker \& D\"oring \cite{becker+doring35}; 
Feder et al.\ \cite{feder+etal66}) or by the related scaled homogeneous nucleation theory (Hale \cite{hale86}).  

The classical homogeneous nucleation theory was developed to describe the nucleation of 
volatile\footnote{A material that readily evaporates.}
materials such as water, hydrocarbons or alcohols at relatively low levels of supersaturation 
(${\cal S} \sim 1.1 - 5.0$) and temperatures ($\sim 300$~K). 
The theory was as such not developed to deal with supersaturated 
refractory\footnote{A material that vaporises only at high temperature.} vapors at
high temperatures.  The theory describes the formation of critical nuclei in a supersaturated
vapor by means of thermodynamic quantities. The essential basic 
assumption of this approach is that the properties of the clusters in the nucleation
regime are given by the extrapolation of the bulk properties
even into the domain of very small clusters or the interpolation of
thermodynamic properties between those of the molecules and the
solid particles. With these assumptions both the thermodynamic
functions such as entropy and enthalpy and the rate coefficients 
describing cluster formation and destruction become simple
analytical functions of the cluster size $N$, which allow a
straightforward calculation of the rate of formation of critical clusters.

A fundamental result of classical nucleation theory is the existence of a
bottleneck for particle formation. The small unstable clusters which form
at random from the gas phase have to grow beyond a certain critical size
$N_{*}$ which corresponds to a maximum in the Gibbs free energy of formation
and separates the domain of small unstable clusters from the large 
thermodynamically stable grains. The rate of grain formation is determined 
by the transition rate ${\cal J}_{*}$ between both regions.
The existence of such a critical cluster size also holds in more realistic 
theories of cluster formation.
However, a review of the available experimental
literature by Nuth \& Ferguson (\cite{nuth+ferguson93}) shows that no experimental data exists
to support the application of classical nucleation theory to the condensation of refractory vapors.
Refractory vapors seem to condense out at different supersaturation 
ratios than volatile materials.

The application of classical homogeneous nucleation theory in astrophysical
discussions of grain formation is sharply criticized by Donn \& Nuth (\cite{donn+nuth85}).
There was some hope that scaled nucleation theory (Hale \cite{hale86}) might be a better way to describe the
condensation of refractory vapors The scaled nucleation theory is a generalization of classical nucleation
theory by scaling the relevant parameters to those of the vapor at the critical temperature
and pressure and the agreement with experimental data for various molecular fluids was rather good 
(Martinez et al.\,\cite{martinez+etal01}). As a result of 
this success scaled nucleation theory was subsequently applied to selected refractory
nucleation data (Hale et al.\ \cite{hale+etal89}) with some success. However, comparison materials 
suggest that refractory materials in general can not be described as accurately 
(i.e. significant deviations occur for lithium, magnesium and bismuth) 
as the molecular 
fluids to which the scaled nucleation theory was originally applied (Nuth \& Ferguson \cite{nuth+ferguson93}). 
It was shown by Martinez et al.\ (\cite{martinez+etal01}) that the reason for the poor agreement of scaled nucleation theory 
to certain refractory materials appear to be, at least in part, the result of using an overestimated
value for the excess surface entropy for liquid metals.  Martinez et al. (\cite{martinez+etal01}) 
conclude that refractory materials, as a class, seem to behave differently than the simple fluids
studied in the original work by Hale (\cite{hale86}) and that the use of bulk liquid properties to describe
a process involving small metallic clusters is problematic and that there therefore is a serious
need for more and better nucleation data for refractory materials.  

A better description of the nucleation of small refractory clusters,
which are needed as input for the moment method,  will most likely have to be guided by 
experiments.  For now we will use the classical
nucleation theory for our model calculations. We are
aware that in this way we introduce uncertainties to our dust description but with no significant
improved theory we find it justified to use classical nucleation
theory as a first crude approximation for calculating the nucleation rate $\cal J_{*}$. 
Once an improved description is derived it will be relatively easy for us to change 
the description of the nucleation rate in the code. 

In the present models the supersaturation ratio $\cal S$ is defined as the ratio
of the partial pressure of carbon atoms in the gas phase divided
by the saturated vapor pressure of solid carbon 
(Eq.\,(\ref{supersaturation})). As shown in Gail \& Sedlmayr (\cite{gail+sedlmayr88})
$\cal S$ depends on the actual lattice temperature of the $N$-cluster.

\begin{figure}
   \centering
   \includegraphics[width=11cm,angle=-90,clip]{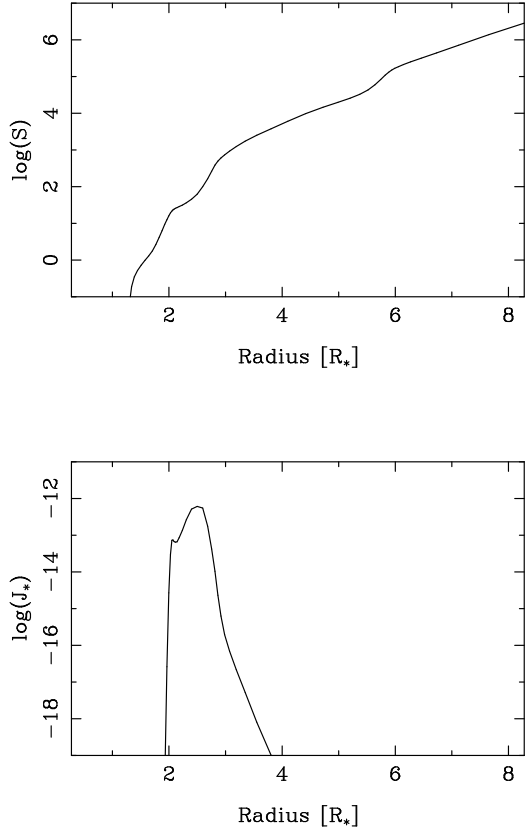}
\caption{A snapshot of the supersaturation $\cal S$ and the nucleation rate 
 ${\cal J}_{\ast}$ for model l13dj10$\rho$199. 
The nucleation rate has a sharp maximum around $3 R_{\odot}$ and at this point
the supersaturation ratio is around 100.}
         \label{supersaturationfig}
   \end{figure}

The value ${\cal S} = 3$ is
adopted following Gail \& Sedlmayr (\cite{gail+sedlmayr87b}) as the 
minimum value for grain nucleation to occur.  As seen from Fig.\,\ref{supersaturationfig},
where the supersaturation ratio ${\cal S}$ and the corresponding nucleation rate ${\cal J}_{\ast}$ 
is shown for one of the calculated models, 
the value of $\cal S$ is on the order of 100 (or at least much larger than unity) 
in the zone where the nucleation rate peaks.

%

\section{Amorphous carbon dust} 

Amorphous carbon particles are considered to be the most common type of dust
present in circumstellar envelopes of carbon-rich AGB stars. The infrared spectra of 
late-type stars generally show a dust emissivity law $Q(\lambda) \sim \lambda^{- \beta}$ with
a spectral index of $\beta \sim 1$ (e.g. Campbell et al.\ \cite{campbell+etal76}; 
Sopka et al.\ \cite{sopka+etal85}; Martin \& Rogers \cite{martin+rogers87};
G\"urtler et al.\ \cite{gurtler+etal96}).  A $\lambda^{-1}$ behavior can be expected in a very
disordered material like amorphous carbon (e.g. Huffmann \cite{huffmann88}; J\"ager et al.\
\cite{jaeger+etal98}). Graphite formation in AGB stars seem unlikely, because of
the absence of the narrow band at $11.52~\mu$m in the observed spectra and the overall shape
of infrared graphite spectra which are  proportional to
$\lambda^{-2}$ (e.g. Draine \& Lee \cite{draine+lee84}). 
This is consistent with physical considerations, predicting the formation
of inhomogeneous grains with crystalline cores surrounded by amorphous
mantles (Gail \& Sedlmayr \cite{gail+sedlmayr84}).

\subsection{Optical properties of dust}

The formation of the dust grains influences the stellar atmosphere in two ways: In the gas
phase chemistry, dust formation results in a depletion of certain elements, which influences the molecular
composition of the gas, and consequently the corresponding opacities. On the other hand, dust grains
have a rather high mass absorption coefficient which often may be comparable to the gas 
opacity or even exceed it. The total opacity of an ensemble of spherical dust grains can be formulated as
\begin{equation}
\kappa(\lambda) = \int_{0}^{\infty} a^2 \, \pi \, {Q_{\rm ext}} \left(a, \lambda \right) \, n \left( a \right) \, da,
\label{kappaext}
\end{equation}
where $n \left( a \right) da$ is the number density of grains in the grain radius interval
between $a$ and $a + da$ and ${Q_{\rm ext}}$ is the extinction efficiency, i.e.
the ratio for the extinction cross section to the geometrical cross section of the grain (Bohren \& Huffman \cite{bohren+huffman83}). This means that the
size distribution function of the dust grains has to be known. However, if the particle size is small compared
to the wavelength of the radiation, the small particle limit (SPL), the extinction efficiency $Q_{\rm ext}$ is given by
\begin{equation}
Q'_{\rm ext}(\lambda) \, = \, \frac{Q_{\rm ext}(\lambda)}{a} \, = \, \frac{8 \pi}{\lambda} \, \Im \left( \frac{1- m^{2}}{2 + m^2} \right),
\end{equation}
where $\Im$ denotes the imaginary part and $m=n+ik$ is the complex refractive index. 
The dependence of the opacity on wavelength and grain size can therefore be separated into two independent factors  
\begin{equation}
\kappa^{\rm SPL}_{\rm ext} \, = \, Q'_{\rm ext}(\lambda) \, \pi \int_{0}^{\infty} a^3 \, n(a) \, da \,
                       = \, \pi \, a_{1}^{3} \, Q'_{\rm ext}(\lambda) \, K_{3}.
\label{kappa}
\end{equation}
This means that the opacity of the dust particles requires the knowledge of the 
moment $K_{3}$ and the monomer radius $a_{1}$, 
\begin{equation}
a_{1} = \left( \frac{3A_{\rm mon} \, m_{p}}{4 \pi \, \rho_{\rm grain}} \right)^{1/3},
\label{monomerradius}
\end{equation}
here $A_{\rm mon}$ is the atomic weight of the monomer (for carbon $A_{\rm mon} = 12.01115$), 
$m_{p}$ the proton mass and $\rho_{\rm grain}$ is the intrinsic density of the condensed grain material. 

In an astrophysical context, knowledge of the moments $K_{i}$ for $i=1,2,3$ is needed if 
the extinction coefficient is required and additionally the moments with $i=4,5,6$ if both absorption 
and scattering are required separately because the scattering coefficient $\kappa_{\rm sca}(\lambda)$
depends on $K_{6}$ (Gail \& Sedlmayr \cite{gail+sedlmayr84}). 

The photospheric spectral energy distribution of AGB stars has its maximum around wavelengths of 1~$\mu$m
(with a sharp decline of the stellar flux toward shorter wavelengths), and both observations and
theoretical arguments indicate that typical grain sizes in these stars are much smaller than 1~$\mu$m. Thus, the limit of
particles being small compared to the wavelength is valid for a major fraction of the spectrum. We therefore
do not need to specify details about the size distribution of the grains to calculate the grain opacities required 
for model atmospheres. However, it may still be necessary to know properties like the size distribution
and shapes of the grains to compute detailed synthetic spectra at very short wavelengths. The complex
refractive index of the material as a function of wavelength can be determined from laboratory measurements.

\subsection{Laboratory measurements of amorphous carbon}

\begin{table*}
\caption{List of the different dust opacities used.}
\begin{center}
\begin{tabular}{|l|c|c|c|c|c|} \hline
Reference & Material & $\rho_{\rm grain}$ & $sp^2$ & Designation & Comments \\
 & name & (g/cm$^{3}$) & \% & in this paper  & \\ \hline
J\"{a}ger et al.\ (\cite{jaeger+etal98}) & cel400  & 1.435 & 67 & {\it J\"{a}ger~400} & completely amorphous \\
J\"{a}ger et al.\ (\cite{jaeger+etal98}) & cel1000 & 1.988 & 80 & {\it J\"{a}ger~1000} & contains graphite (2~nm) crystallites \\
Rouleau \& Martin (\cite{rouleau+martin91}) & AC2 & 1.85 & -  & {\it Rouleau} & \\ \hline
\end{tabular}
\end{center}
\label{opacitydust}
\end{table*}

Due to the nature of amorphous carbon dust materials they span a broad range of micro-physical
properties. In the models presented
here we have used three different laboratory measurements (referred to as {\it J\"ager~1000, J\"ager~400} and
{\it Rouleau}, see Table \ref{opacitydust}) to describe the opacity of the dust grains formed
in the circumstellar envelope.

The data by J\"ager et al. (\cite{jaeger+etal98}) are produced by pyrolizing cellulose materials at
different temperatures. The materials are characterize in exemplary detail. In this study we have used the data
synthesized at 400$^{\circ}$\,C  and 1000$^{\circ}$\,C. The two materials differ in the bonding, where
the {\it J\"ager400} material has more single bonds (the carbon atoms are mainly $sp^3$ hybridized) while the
{\it J\"ager1000} material has more double bonds (the carbon atoms are mainly $sp^2$ hybridized).
Reflectance spectra were obtained of the samples and from these the complex refractive index ($m$)
was derived by the Lorentz oscillator method (see e.g.\ Bohren \& Huffman \cite{bohren+huffman83}, Chap.\ 9).

Rouleau \& Martin (\cite{rouleau+martin91}) produced synthetic optical constants ($n$ and $k$; 
$m = n + ik$) based on measurement of sub-micron amorphous carbon particles by Bussoletti et al.
(\cite{bussoletti+etal87}). The particles were produced by striking an arc between two amorphous carbon 
electrodes in a controlled Ar atmosphere.

\subsection{Grain equilibrium temperature}
\label{slobeSec}

The presence of dust grains influences both the momentum and the energy balance of
the atmosphere. We assume
complete momentum coupling of gas and dust, which means that the momentum
gained by the dust from the radiation field is directly transferred to the gas.
On the other hand, the transfer of internal energy between gas and dust is
negligible compared to the interaction of each component with the radiation field 
(cf.\ Gauger et al.\ \cite{gauger+etal90}).  We therefore assume that the grain temperature 
is given by the condition of radiative equilibrium
\begin{equation}
\chi_{J} \, J - \chi_{S} \, S_{\rm d} = 0 \hspace{0.2 cm} \Rightarrow \hspace{0.2 cm} T_{\rm d} = \left( \frac{\chi_{J}}{\chi_{S}} \right)^{\frac{1}{4}} T_{\rm r}
\end{equation}
where the source function is equal to the Planck function $S_{\rm d} = B(T_{\rm d})$. The frequency-integrated dust opacities are defined by
\begin{eqnarray}
\chi_{J} &=& \frac{\int_{\nu} \chi_{\nu} \, J_{\nu} \, d\nu}{\int J_{\nu} \, d\nu} \\
\chi_{S} &=& \frac{\int_{\nu} \chi_{\nu} \, S_{\nu} \, d\nu}{\int S_{\nu} \, d\nu} 
    = \frac{\int_{\nu} \chi_{\nu} \, B_{\nu}(T_{\rm d}) \, d\nu}{\int B_{\nu}(T_{\rm d}) \, d\nu} \\ \nonumber
    &=& \frac{\int_{\nu} \chi_{\nu} \, B_{\nu}(T_{\rm d}) \, d\nu}{B(T_{\rm d})}
\end{eqnarray}
where $J_{\nu}$ is the radiation energy density, 
and the radiation temperature $T_{\rm r} = (J \pi/\sigma)^{1/4}$.

\begin{figure}
   \centering
    \includegraphics[width=9cm,angle=0,clip]{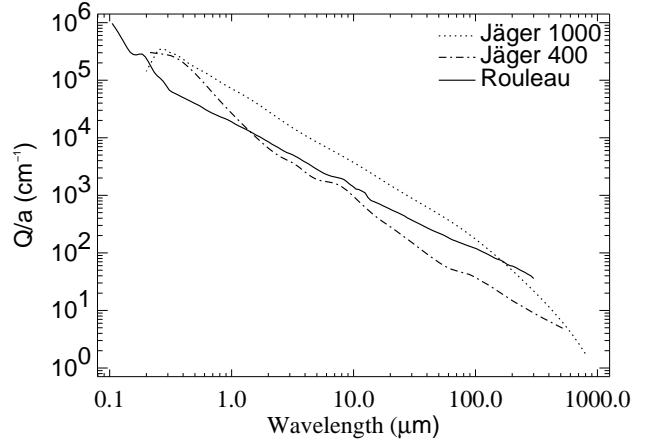}
\caption{Wavelength dependence of the dust absorption efficiency $Q'_{\rm ext}$
           (see Table \ref{opacitydust} for dust annotation).}
\label{gray}
\end{figure}

For a gray opacity we would obtain $\chi_{J} \equiv \chi_{S}$ for the opacities and 
the temperatures $T_{\rm d} \equiv T_{\rm r}$ as in previous models.
Using frequency-dependent radiative transfer the opacities $\chi_{J}$ and $\chi_{S}$ differ. 
For the data shown in Fig.\,\ref{gray}, the absorption coefficients decrease with increasing
wavelength in the infrared region of the spectrum, leading to $T_{\rm d} > T_{\rm r}$
for all data sets shown. The difference between $T_{\rm d}$ and $T_{\rm r}$ becomes larger with 
an increasing slope of $Q'_{\rm ext}(\lambda)$ and the difference may reach several $10^{2}$~K
in the dust formation zone (H\"ofner et al.\ \cite{hoefner+etal02}).
In other words, the steeper the dependence on wavelength, the larger the difference between the
equilibrium grain temperature and the radiation temperature. Therefore,
the steeper slope of {\it J\"ager~400} data results in a relatively high grain temperature compared to 
{\it J\"ager~1000} and {\it Rouleau}.


\section{Results}

%

One of the  main reasons for the huge mass-loss of AGB stars seems to be the presence of
newly formed dust grains. The strong shock waves in the stellar atmosphere cause a
levitation of the outer layers. The cool and relatively dense environment which results
from the levitation provides favorable conditions for the formation of molecules and
grains.  Due to its high opacity and the resulting radiative pressure,
the dust has a strong influence on the structure of the atmosphere and the wind properties.


\subsection{Influence of the dust extinction coefficient}

%
   \begin{figure}
   \centering
    \includegraphics[angle=0,width=9cm,clip]{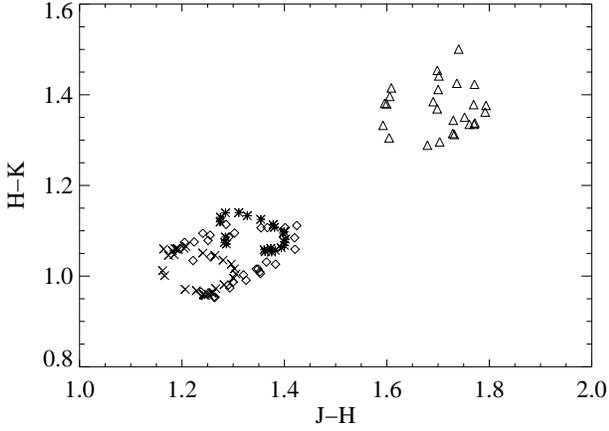}
     \caption{The colors (J$-$H) vs. (H$-$K) of the four  
     models l13dj10$\rho$199 ($\diamond$),  
     l10dj10$\rho$199 ($\triangle$), 
     l13drou$\rho$185 ($\times$) and 
      l10drou$\rho$185 ($\ast$). 
     There are two differences between the two sets of models.
     (1) The assumed value of the luminosity ($L_{\star}$) and corresponding
   effective temperature ($T_{\star}$), and (2) the dust opacity data used 
     ({\it J\"ager~1000} and {\it Rouleau}). All other model values are kept 
      constant. See Table \ref{cool_warm_table} for details.}
         \label{cool_warm}
   \end{figure}

The influence of the dust extinction coefficient on 
the winds of the dynamical models is shown in 
Table\,\ref{cool_warm_table} and Fig.\,\ref{cool_warm}.
The models depend on the choice of laboratory measurements of
$Q'_{\rm ext}(\lambda)$ as already shown 
for gray models in Andersen et al.\ (\cite{andersen+etal99}).
In contrast to the gray models, 
for the frequency-dependent models
both the absolute value and the slope of the dust opacity data 
as a function of $\lambda$ become 
relevant. The absolute value of the grain opacities
mainly affects the terminal velocity of
the winds, while the slope has significant
influence on the grain temperature as discussed in Sect.\,\ref{slobeSec}.
In the models where the {\it J\"ager~400} data is used for the opacity 
of the grains, the slope of the extinction efficiency as a function of $\lambda$
dictates a high grain temperature which prevents dust formation for the chosen stellar parameters.
Models with identical stellar parameters using {\it J\"ager~1000} and
{\it Rouleau} data on the other hand develop dust-driven winds.

Figure \ref{cool_warm} and Table \ref{cool_warm_table} show that 
it is not possible to distinguish from near infrared photometry
alone between the different models, since the hotter model using the {\it J\"ager~1000}
opacity data has similar colors as the cooler model using
the optical properties from the {\it Rouleau} data. 
This is a consequence of the fact that these two models have circumstellar envelopes with
comparable optical depth at these wavelengths as can be demonstrated by the following rough estimate; 
the optical depth of a given layer $d\tau$ can be expressed in terms of the extinction coefficent, the local degree of condensation $f_{\rm c}$,
mass-loss rate and flow velocity by
\begin{eqnarray}
d\tau = \kappa \, dr \, &=& \, \pi \, a^{3}_{1} \, Q'_{\rm ext}(\lambda) \, K_{3}(r) \, dr \\ \nonumber & \propto & \, \, a^{3}_{1} \, Q'_{\rm ext} \, 
        f_{\rm c} \, \rho_{\rm gas} \, dr \, \\ \nonumber
         &\propto& \, \, a^{3}_{1} \, Q'_{\rm ext} \, f_{\rm c} \, \frac{\dot{M}}{u} \, \frac{1}{r^2} \, dr
\label{tau-kappa}
\end{eqnarray}
where we have used Eq.\,(\ref{kappa}), $K_{3} \propto f_{\rm c} \, \rho_{\rm gas}$ and $\rho_{\rm gas} = \dot{M} / 4\pi r^2 u$ 
to replace the gas density $\rho_{\rm gas}$.
To estimate the relative optical depths of the circumstellar envelopes of different models
we assume that we can replace $f_{\rm c} \dot{M}/u$ by the the time-averaged values 
${\langle f_{\rm c} \rangle} \dot{M}/{\langle u \rangle}$ at the outer boundary.
For models l13dj10$\rho$199 and l10drou$\rho$185 the differences in mass-loss rate, degree of 
condensation and outflow velocity as well as the extinction efficiency and the monomer radius
compensate each other in such a way that the quantity $a^{3}_{1} Q'_{\rm ext} {\langle f_{\rm c} \rangle} \dot{M}/{\langle u \rangle}$
is comparable for both models while the value for l10dj10$\rho$199 differs by about a factor of two.
Although it is impossible to distinguish the models l13dj10$\rho$199 and l10drou$\rho$185 by their near infrared colors alone
it would be possible to do this by a combination of photometry and high-resolution spectroscopy
due to their significantly different outflow velocities (of about a factor of four).

\begin{table*}
\caption{Influence of the extinction efficiency of the dust.
         {\it Model parameters:} $M_{\star} = 1.0 \,M_{\odot}$,
         sticking coefficient $\alpha_{\rm C} = 0.37$, $\alpha_{\rm C_{2}} = 0.34$, 
         $\alpha_{\rm C_{2}H} = 0.34$ and $\alpha_{\rm C_{2}H_{2}} = 0.34$,
         ${\varepsilon_{\rm C}/\varepsilon_{\rm O}} = 1.4$,
         period $P = 650$\,d, piston velocity ${\Delta u_{\rm p}} = 4$\,km/s,
         luminosity $L_{\star}$, 
          temperature $T_{\star}$,
          radius $R_{\star}$, 
         dust opacity data $\kappa_{\rm dust}$, intrinsic dust density $\rho_{\rm grain}$;
         {\it Results:} Mass loss rate $\dot{M}$,
         mean velocity at the outer boundary ${\langle u \rangle}$,
         mean degree of condensation at the outer boundary ${\langle f_{\rm c} \rangle}$.}
\begin{center}
 \begin{tabular}{|l|c|c|c|c|c||c|c|c|c|}
 \hline
Model& $L_{\star}$ & $T_{\star}$ & $R_{\star}$  & $\kappa_{\rm dust}$& $\rho_{\rm grain}$ & $\dot{M}$ & ${\langle u \rangle}$ & ${\langle {f_{\rm c}} \rangle}$ & Symbol in \\ 
 & [$L_{\odot}$] & [K] & [${\rm {R_{\odot}}}$] & & [g/cm$^{3}$] & [${\rm {M_{\odot}} / yr}$] & [km/s] & & Fig.\,\ref{cool_warm} \\ \hline
l13dj10$\rho$199&  13000 & 2700 & 521 & {\it J\"ager~1000} & 1.99 & $5.6 \cdot 10^{-6}$ & 15 & 0.05 &  $\diamond$  \\
l13drou$\rho$185& 13000 & 2700 & 521 & {\it Rouleau} & 1.85 &  $4.9 \cdot 10^{-6}$ & 7.4 & 0.10 &  $\times$  \\
l13dj04$\rho$144& 13000 & 2700 & 521 & {\it J\"ager~400} & 1.44 &  - & - & - & -  \\
l10dj10$\rho$199& 10000 & 2600 & 493 & {\it J\"ager~1000} & 1.99 &  $7.0 \cdot 10^{-6}$ & 16 & 0.10 & $\triangle$ \\
l10drou$\rho$185&  10000 & 2600 & 493 & {\it Rouleau} & 1.85 &  $2.3 \cdot 10^{-6}$ & 3.6 & 0.12 & $\ast$ \\
l10dj04$\rho$144&  10000 & 2600 & 493 & {\it J\"ager~400} & 1.44 &   - & - & - & -  \\
\hline
\end{tabular}
\end{center}
\label{cool_warm_table}
\end{table*}
\begin{table*}
\caption{Influence of the intrinsic density of the dust.
         {\it Model parameters:} Same as Table\,\ref{cool_warm_table}.}
\begin{center}
\begin{tabular}{|l|c|c|c|c||c|c|c|c|}
\hline
Model& $L_{\star}$  & $T_{\star}$ & $\kappa_{\rm dust}$&  $\rho_{\rm grain}$      & $\dot{M}$ & ${\langle u \rangle}$ & ${\langle {f_{\rm c}} \rangle}$ & Symbol in \\
     &[$L_{\odot}$] & [K]         & g/cm$^{3}$         &[${\rm {M_{\odot}} / yr}$]& [km/s] &       &                   &  Fig.\,\ref{rho} \\
\hline
l13dj10$\rho$199& 13000 & 2700& {\it J\"ager~1000} & 1.99 & $5.6 \cdot 10^{-6}$ & 15 & 0.05 & $\diamond$  \\
l13dj10$\rho$225&13000 & 2700& {\it J\"ager1000} & 2.25 & $7.3 \cdot 10^{-6}$ & 21 & 0.11 & $\ast$  \\
l13drou$\rho$185&13000 & 2700& {\it  Rouleau} & 1.85 &  $4.9 \cdot 10^{-6}$ & 7.4 & 0.10 & $\times$ \\
l13drou$\rho$225& 13000 & 2700& {\it Rouleau} & 2.25 &  $8.2 \cdot 10^{-6}$ & 18 & 0.31 & $\triangle$  \\
l13dj04$\rho$144&13000 & 2700& {\it J\"ager400} & 1.44  & - & - & - & -  \\
l13dj04$\rho$225& 13000 & 2700& {\it J\"ager400} & 2.25  & $2.1 \cdot 10^{-8}$ &  1.38 & 0.13 & -  \\
\hline
\end{tabular}
\end{center}
\label{density}
\end{table*}
\begin{table*}
\caption{Influence of the sticking coefficient for the dust formation.
         {\it Model parameters:} Same as Table\,\ref{cool_warm_table}
         except for the sticking coefficients: $\alpha_{\rm C}$, $\alpha_{\rm C_{2}}$, 
         $\alpha_{\rm C_{2}H}$ and $\alpha_{\rm C_{2}H_{2}}$, dust opacity
         $\kappa_{\rm dust}$ = {\it Rouleau}, and the intrinsic dust
        density $\rho_{\rm grain} = 1.85$~g/cm$^3$ for all models.}
\begin{center}
 \begin{tabular}{|l|c|c|c|c|c|c||c|c|c|c|}
 \hline
Model& $L_{\star}$ & $T_{\star}$ & $\alpha_{\rm C}$ & $\alpha_{\rm C_{2}}$ & $\alpha_{\rm C_{2}H}$ & $\alpha_{\rm C_{2}H_{2}}$ & $\dot{M}$ & ${\langle u \rangle}$ & ${\langle {f_{\rm c}} \rangle}$ & Symbol in \\
     &[$L_{\odot}$]& [K]         &              &                  &                   &                       & [${\rm {M_{\odot}} / yr}$] & [km/s]                &                                 & Fig.\,\ref{stickcolor} \\
 \hline
l13drou$\rho$185& 13000 & 2700 & 0.37 & 0.34 & 0.34  & 0.34   & $4.9 \cdot 10^{-6}$ & 7.4 & 0.10 & $\ast$  \\
l13drou$\rho$185$\alpha$02&  13000 & 2700 & 0.20 & 0.20 & 0.20 & 0.20 & $3.4 \cdot 10^{-6}$ & 3.9 & 0.09 & - \\
l13drou$\rho$185$\alpha$05& 13000 & 2700 &  0.50 & 0.50 & 0.50 & 0.50 & $5.8 \cdot 10^{-6}$ & 11 & 0.12 & - \\
l13drou$\rho$185$\alpha$10& 13000 & 2700 &  1.00 & 1.00 & 1.00 & 1.00 & $7.0 \cdot 10^{-6}$ & 17 & 0.22 & $\triangle$ \\ 
l10drou$\rho$185&  10000 & 2600 &  0.37 & 0.34 & 0.34 & 0.34 & $2.3 \cdot 10^{-6}$ & 3.6 & 0.12 & - \\
l10drou$\rho$185$\alpha$10&  10000 & 2600 & 1.00 & 1.00 & 1.00 & 1.00 & $7.0 \cdot 10^{-6}$ & 12 & 0.23 & - \\
\hline
\end{tabular}
\end{center}
\label{sticktable}
\end{table*}
%


\subsection{Influence of the intrinsic dust density}

%
   \begin{figure}
   \centering
     \includegraphics[angle=0,width=9cm,clip]{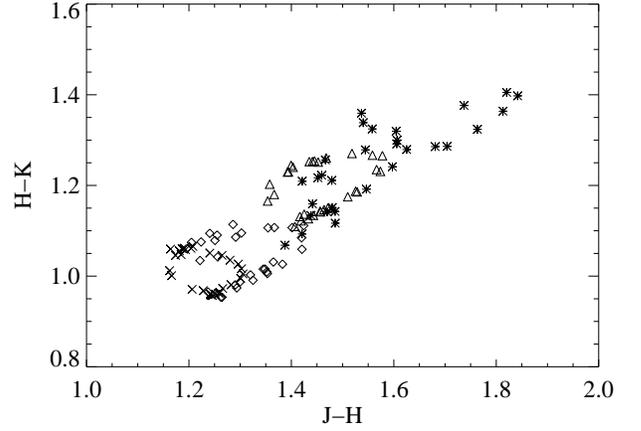}
     \caption{The colors (J$-$H) vs. (H$-$K) for the four models 
    l13dj10$\rho$199 ($\diamond$), l13dj10$\rho$225 ($\ast$),
    l13drou$\rho$185 ($\times$) and l13drou$\rho$225 ($\triangle$). 
     We see the results for two different opacity data ({\it J\"ager~1000}
     and {\it Rouleau}). Two different models for each opacity data are shown.
      The difference between the two models with the same opacity  is the 
     assumed value of the intrinsic density ($\rho_{\rm grain}$) of the dust 
       material formed in the model. The values correspond to the one 
    determined in the laboratory for the material and the value for 
graphite (2.25 g/cm$^3$) which was used in previous models. See Table \ref{density}
    for details.}
         \label{rho}
   \end{figure}
%

When the moment method was 
developed by Gail \& Sedlmayr (\cite{gail+sedlmayr88}) the intrinsic
density for the amorphous carbon material used in that generation of models 
(Maron \cite{maron90}) was not known. The value of graphite (2.25 g/cm$^{3}$)
was therefore assumed. This value was later used in most 
existing models based on the moment method (e.g.\ Fleischer et al.\ \cite{fleischer+etal92};
H\"ofner \& Dorfi \cite{hoefner+dorfi97}).

In the previous section we have described models which use 
optical properties represented by three different amorphous carbon
materials for which the intrinsic densities
have been measured in the laboratory. Here, we compare them with three models
where the same optical properties were used but where we assumed the higher 
value for the intrinsic density of graphite instead of the measured value 
of the material (while keeping all other parameters constant).

The result of using the higher density of graphite instead of the correct values
can be seen in Table \ref{density} and Fig.\,\ref{rho}. Figure \ref{rho} demonstrate   
that the models become much redder when the intrinsic density of graphite
is used instead of the respective values for amorphous carbon.
The large increase in the mean degree of condensation 
${\langle f_{\rm c} \rangle}$ 
at the outer boundary  implies that more dust is formed. This 
results in an increased radiation pressure on the dust grains and the models
therefore show a higher mean velocity
${\langle u \rangle}$ at the outer boundary.
At the same time the mass-loss rates increases so that 
$a^{3}_{1} {\langle f_{\rm c} \rangle} \dot{M}/{\langle u \rangle}$,
which is a measure of the optical depth of the envelope,
is significantly higher for both models using the density of graphite
compared to the respective models with the consistent dust densities.

Even a small
increase of about 10\% in the density of the dust material (as it is the case from
model l13dj10$\rho$199 to l13dj10$\rho$225) 
results in a doubling of the degree of condensation and a
substantial increase of the outflow velocity ${\langle u \rangle}$ and the mass-loss rate.  
For the models l13drou$\rho$185 and l13drou$\rho$225 where
the difference in the value used for the intrinsic dust
density is about 20\%, the estimated mass-loss rates differ by almost a factor of two.
For the models using the {\it J\"ager~400} material (l13dj04$\rho$144 and l13dj04$\rho$225),
where the difference is almost 40\%, using the measured material value instead of  
the higher value for graphite results in a model that will not develop a wind at all.

The intrinsic dust density of the dust material  $\rho_{\rm grain}$ is so significant
for the obtained results because when it 
is increased the monomer radius $a_{1}$ decreases (see Eq.\,(\ref{monomerradius})). 
This influences both the grain growth and the dust opacity.
The net growth rate $1/\tau$
and the opacity $\kappa$ dependent on the monomer radius as
$1/\tau \propto a_{1}^{2}$ (through the monomer surface $A_{1}$) and $\kappa \propto a_{1}^{3}$, respectively. 
A reduction of $a_{1}$ decreases the dust mass absorption coefficient 
stronger than the growth rate, leading both to a slower acceleration and a more
efficient growth of the grains. The opacity both depends on $a_{1}^3$ (which decrease slightly)
and the degree of condensation $f_{\rm c}$ (which increases) with the later effect dominating,
which leads to a higher final outflow velocity ${\langle u \rangle}$ of the wind, despite a reduction of the dust
absorption per gram of dust.


\subsection{Influence of the estimated sticking coefficient}

%
   \begin{figure}
   \centering
    \includegraphics[angle=0,width=9cm,clip]{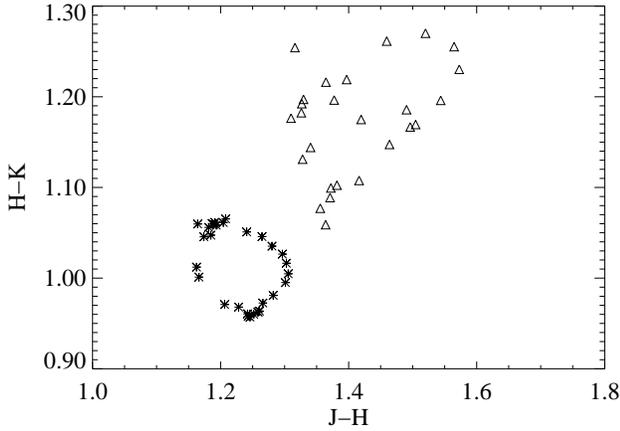}
     \caption{The colors (J$-$H) vs. (H$-$K) for the two models 
    l13drou$\rho$185 ($\ast$) and l13drou$\rho$185$\alpha10$ ($\triangle$).
      The only difference between the two models is the assumed value of 
      the sticking coefficient. The larger the sticking coefficient is
      assumed to be, the more efficient will the grain formation  be and
      as a consequence the models will look redder when observed in
      (H$-$L), (J$-$H) or (K$-$L). See Table \ref{sticktable} for details.}
         \label{stickcolor}
   \end{figure}
%

The sticking coefficient (also called the reaction efficiency
factor) $\alpha$ enters into the net growth rate of the
dust grains (Eq.\,(\ref{tau})). However, $\alpha$
is not definitely known as long as we do
not know explicitly the sequence of chemical reactions
responsible for the dust formation.
To demonstrate how uncertainties in the
value of $\alpha$ will influence the results of the
models we have varied this parameter in otherwise identical
models. 

In the case of carbon dust formation the most important growth species
is expected to be C$_{2}$H$_{2}$, but also C, C$_{2}$ and C$_{2}$H
contribute to the growth (Gail \& Sedlmayr \cite{gail+sedlmayr88}).
In our earlier models we have used the values $\alpha_{\rm C} = 0.37$,
$\alpha_{\rm C_{2}} = 0.34$, $\alpha_{\rm C_{2}H} = 0.34$ and 
$\alpha_{\rm C_{2}H_{2}} = 0.34$ to describe the grain growth. 

Gail \& Sedlmayr (\cite{gail+sedlmayr84}) used
the value $\alpha_{\rm C} = 0.3$ adopted from 
Landolt-B\"ornstein (\cite{landolt-bornstein68}).
Later Gail \& Sedlmayr (\cite{gail+sedlmayr88}) argued 
that the  sticking coefficient $\alpha$ must be on the order
of unity, because it is expected that neutral radical reactions
play a dominant role in the formation process of carbon grains.
However, if substantial energy barriers are involved
in the reactions $\alpha$ may well be less by several
orders of magnitude. It is important to remember
that if $\alpha$ is small the assumption that 
nucleation can be treated as a time-independent
process will no longer hold. Small values
of $\alpha$  should be accompanied by a 
time-dependent treatment of the dust nucleation
(Gail \& Sedlmayr \cite{gail+sedlmayr88}).

Salpeter (\cite{salpeter73}) has shown that the latent heat released when a
monomer attaches itself to an $N$-mer may lead to a small sticking
probability if $N$ is small. Thus the sticking coefficient $\alpha$ 
for clusters with $N \approx N_{*}$ may be considerably smaller than the
sticking coefficient for clusters with $N \gg N_{*}$. The sticking coefficient
is of the order unity for bulk material (Pound \cite{pound72}). 

To test the assumptions for the sticking
coefficient we have calculated four models for a
star with $L_{\star} = 13000~L_{\odot}$, 
$T_{\star} = 2700$~K and with the sticking coefficient ($\alpha$) varying
from $0.2 - 1.0$ and two cooler models with
$L_{\star} = 10000~L_{\odot}$ and $T_{\star} = 2600$~K.

\begin{figure*}
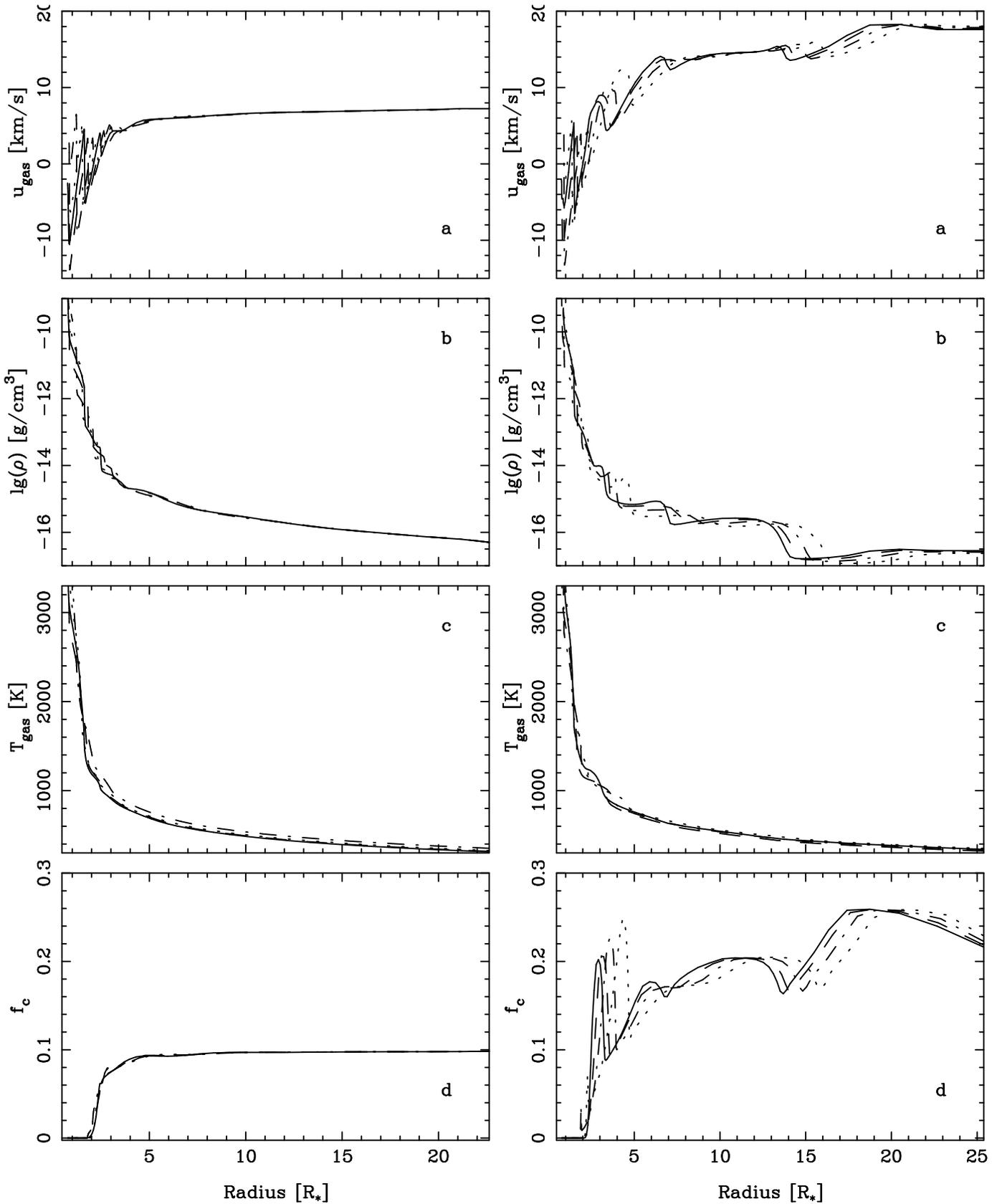

   \centering
  \vspace{23 cm}
   \includegraphics{H4051_F6.ps}
   \includegraphics{H4051_F7.ps}
\caption{Models l13drou$\rho$185 (left) and 
         l13drou$\rho$185$\alpha$10 (right). For four different
         phases of the two models (a) the
         outflow velocity of the gas, (b) the gas density,
         (c) the gas temperature and (d) the degree of
         condensation is shown. The only
         difference between the two models is the value used 
         for the sticking coefficient. It is evident that
         the choice of sticking coefficient has a significant
         influence on the predicted outflow velocity 
         and the degree of condensation.}
\label{stick}
\end{figure*}

It is clear from Table\,\ref{sticktable} and Fig.\,\ref{stickcolor} 
(showing how the colors of two models depend on the chosen value
for the sticking coefficient) that
using the value of $1$ as suggested by Gail \& Sedlmayr (\cite{gail+sedlmayr88})
compared to our previous values of $0.34-0.37$
results in a noticeable reddening of the colors of the stellar model.
Fig.\,\ref{stick} 
 shows the mean outflow velocity, the gas density, the gas temperature and 
the mean degree of condensation for four different phases of the two
different models l13drou$\rho$185$\alpha$ and l13drou$\rho$185$\alpha$10. 
The higher value of the sticking coefficient 
increases both the degree of condensation and the outflow velocity by
about a factor of two. In addition the mass-loss rate increases, leading
to a higher optical depth of the circumstellar envelope.

From Fig.\,\ref{stick} it can be seen that dust formation occurs beyond $\approx
2 R_{\star}$ independent of the value assumed for the sticking coefficient
since the onset of condensation is determined mainly by the temperature.  With the low
choice of sticking coefficient only about 10\% of the condensable
carbon material present in the gas actually condenses into grains, while for the
much more favorable choice of sticking coefficient ($\alpha=1$) twice as much
material condenses into grains. A complete condensation of carbon grains is
prevented by the rapid velocity increase after the onset of avalanche nucleation
and grain growth, and the subsequent rapid dilution of the gas. 

%

\subsection{Influence of the surface tension for the dust} 

In classical homogeneous nucleation theory, the surface tension of the
grain material is used to describe the gain of enthalpy 
by forming a grain out of $N$ monomers in the gas phase.

The surface tension $\sigma_{\rm grain}$ of amorphous carbon is not known from 
laboratory experiments, it has therefore become custom to use 
values for graphite, e.g.\ one of the values given by 
Tabak et al.\ (\cite{tabak+etal75}).
One problem however is that there are huge variations in the
values determined due to the anisotropy of graphite.

As already demonstrated by Tabak et al.\ (\cite{tabak+etal75}),
varying the value of the surface tension may produce an
enormous change in the nucleation rate. To determine how significant
the prescribed value for the surface tension 
is for our results we have varied it around the
value of 1400~erg/cm$^{2}$ which was used in previous models
(see Table\,\ref{surface}). 

\begin{table}
\caption{Influence of the surface tension for the dust formation. 
         {\it Model parameters:} Same as for model l13drou$\rho$185 in 
         Table\,\ref{cool_warm_table}
         but with different values for the surface tension $\sigma_{\rm grain}$. }
\begin{center}
 \begin{tabular}{|l|c||c|c|c|}
 \hline
Model& $\sigma_{\rm grain}$ & $\dot{M}$ & ${\langle u \rangle}$ & ${\langle {f_{\rm c}} \rangle}$  \\
  &   [erg/cm$^{2}$] & [${\rm {M_{\odot}} / yr}$] & [km/s] &  \\
 \hline
l13drou$\rho$185$\sigma$10& 1000 & $8.5 \cdot 10^{-6}$ & 35 & 0.76 \\
l13drou$\rho$185 & 1400 & $4.9 \cdot 10^{-6}$ & 7.4 & 0.10  \\
l13drou$\rho$185$\sigma$18& 1800 & - &  - & - \\
\hline
\end{tabular}
\end{center}
\label{surface}
\end{table}

For the particular model, altering the value of the 
surface tension of the dust grains by 28\% around the value 
of graphite will make the difference between obtaining mass-loss 
or not. The value of the surface tension also has a substantial influence on how 
much of the available material in the circumstellar envelope 
will condense into dust grains. For comparison, the measured surface
tension for other materials are: Fe ($\sigma_{\rm grain} = 1400$~erg/cm$^{2}$), MgS 
($\sigma_{\rm grain} = 800$~erg/cm$^{2}$)  and SiO ($\sigma_{\rm grain} = 500$~erg/cm$^{2}$),
see Gail \& Sedlmayr (\cite{gail+sedlmayr86}).


\subsection{Comparison with observations}

\begin{table*}
\caption{Data for three observed carbon-rich Mira's with similar colors and 
outflow velocities as the calculated models (Tables \ref{cool_warm_table} 
and \ref{sticktable}). The observed velocities and mass-loss rates are from Olofsson et al.\ 
(\cite{olofsson+etal93}).}.
\begin{center}
 \begin{tabular}{|l|c|c|c|c|c|c|}
 \hline
Star  & Period  & $v_{\rm out}$ & $\dot{M}_{\rm gas}$ & $\dot{M}_{\rm dust}^{lower}$ & $\dot{M}_{\rm dust}^{upper}$  &  Symbol in   \\ 
 & [days] & [km/s] & [${\rm {M_{\odot}} / yr}$] & [${\rm {M_{\odot}} / yr}$]& [${\rm {M_{\odot}} / yr}$] & Fig.\,\ref{massloss_color} \\
\hline
R For & 389 & 16.3 & $1.0 \times 10^{-6}$ & $2.4 \times 10^{-9}$ & $3.5 \times 10^{-9}$ &  $\ast$ \\
R Lep & 427 & 17.0 & $7.0 \times 10^{-7}$ & $1.9 \times 10^{-9}$ & $2.8 \times 10^{-9}$ &  $\times$ \\
R Vol & 454 & 18.5 & $2.5 \times 10^{-6}$ & $4.2 \times 10^{-9}$ & $5.1 \times 10^{-9}$ &  $+$ \\  
    & & & & & & \\ \hline
Model  & Period  & ${\langle u \rangle}$ & $\dot{M}_{\rm gas}$ & \multicolumn{2}{c|}{$\dot{M}_{\rm dust}$} &  Symbol in   \\ 
 & [days] & [km/s] & [${\rm {M_{\odot}} / yr}$] & \multicolumn{2}{c|}{[${\rm {M_{\odot}} / yr}$]} & Fig.\,\ref{massloss_color} \\ \hline
l13dj10$\rho$199 & 650 & 15 & $5.6 \times 10^{-6}$ & \multicolumn{2}{c|}{$6.3 \times 10^{-10}$} &  $\triangle$ \\ 
l13drou$\rho$185$\alpha$10 & 650 & 17 & $7.0 \times 10^{-6}$ & \multicolumn{2}{c|}{$3.5 \times 10^{-9}$} &  $\Box$ \\
l10dj10$\rho$199 & 650 & 16 & $7.0 \times 10^{-6}$ & \multicolumn{2}{c|}{$3.5 \times 10^{-9}$}  &  $\bigcirc$ \\ \hline
\end{tabular}
\end{center}
\label{massloss}
\end{table*}

%
   \begin{figure}
   \centering
    \includegraphics[angle=0,width=9cm,clip]{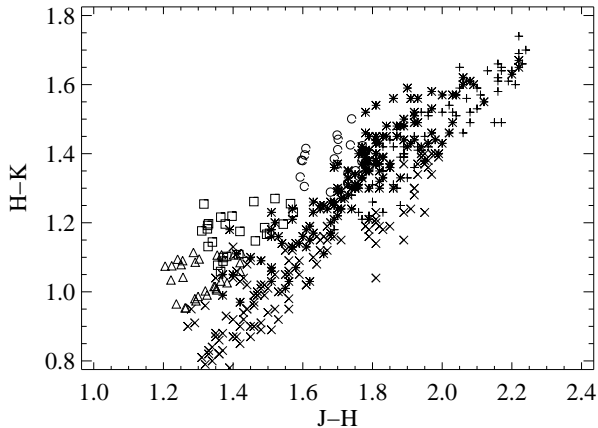}
     \caption{The colors ($J-H$) vs. ($H-K$) for the three models
       l13dj10$\rho$199 ($\triangle$), l13drou$\rho$185$\alpha$10 ($\Box$) 
       and l10dj10$\rho$199 ($\bigcirc$), and three comparable observed carbon Miras 
        R Vol ($+$), R For ($\ast$),
        R Lep ($\times$) (observations from Whitelock  et al.\ \cite{whitelock+etal97}).}
         \label{massloss_color}
   \end{figure}
%

Near-infrared colors including JHKL\footnote{Central filter wavelength
[$\mu$m]: J = 1.22, H = 1.63, \\ K = 2.19, L = 3.45.} magnitudes were
calculated for a selection of the models.  The filter zero points were
calculated from a Vega model of Dreiling \& Bell
(\cite{dreiling+bell80}), under the assumption that the Vega model has
0.0 mag in all filters.

Whitelock et al.\ (\cite{whitelock+etal97}) present JHKL light curves
for 11 large-amplitude carbon variables. For a comparison between models
and observations we picked the carbon Miras R For, R Lep and R Vol.
These stars have moderately thick dust shells and colors
comparable with our models (see Fig.\,\ref{massloss_color}). According to
Olofsson et al.\ (\cite{olofsson+etal93}) these stars have outflow
velocities of about $16 - 19$~km/s. From our set of models we selected
those with comparable velocities (see Table \ref{massloss}). We
preferred outflow velocities over other wind properties when selecting
the models for this comparison, since this is the most directly
observable quantity. The colors on the other hand contain entangled
information about the mass-loss rates, the ouflow velocities, the degree
of condensation, and the optical properties of the grains.

We have not selected the stars for their derived stellar parameters but
for their observable wind properties.
The observed stars (R For, R Lep and R Vol) and the models
(l13drou$\rho$185$\alpha$10 and l10dj10$\rho$199) have similar dust mass
loss rates and near infrared colors while the gas mass-loss rates
for the observed stars and the models differ significantly (see  Table
\ref{massloss}). We preferred the dust mass-loss rates over the gas
mass-loss as a criterion for the comparison since the near IR colors
are strongly affected by the dust opacity, as demonstrated by the
figures in the preceeding section.

The dust mass-loss rates of the observed stars were estimated using the
$60~\mu$m IRAS flux. In general
there is expected to be a relation between the dust mass-loss rate and
the strength of the dust emission. Following Sopka et al.\
(\cite{sopka+etal85})
Olofsson et al.\ (\cite{olofsson+etal93}) 
derive the dust mass-loss rates of these stars using
\begin{equation}
\dot{M}_{\rm dust} = 2.2 \times 10^{-15} S_{60} v_{d} D^{2} L^{-0.5}
\,\,M_{\odot} {\rm yr}^{-1},
\end{equation}
where $S_{60}$ is the IRAS $60 \mu$m flux in Janskys, $v_{d}$ is the
dust expansion velocity in
km/s, $D$ is the distance to the source in parsecs and $L$ is the
luminosity in units of $L_{\odot}$.
For the dust expansion velocity Olofsson et al.\
(\cite{olofsson+etal93}) use two different values, one where the upper
limit to the drift velocity is obtained by considering the radiation
force on the grains and the drag force due to gas-grain collisions in
the limit of supersonic motion. The lower limit for the dust expansion
velocity is derived by neglecting drift. As a consequence two different
dust mass-loss rates are given by Olofsson et al.\
(\cite{olofsson+etal93}), one for the upper limit $\dot{M}_{\rm
dust}^{upper}$ and one for the lower limit $\dot{M}_{\rm dust}^{lower}$.

The gas mass-loss rates of the observed stars given in Table
\ref{massloss} were derived by Olofsson et al.\
(\cite{olofsson+etal93}) based on observed CO emission using equation
(5) of Knapp \& Morris (\cite{knapp+morris85}) which has the form
\begin{equation}
\dot{M}_{\rm gas} = \left( 1 + \frac{4n_{\rm He}}{n_{\rm H}} \right)
\frac{T_{\rm mb} v_{e}^{2} D^{2}}{A(B,J) f_{\rm CO}^{0.85}} M_{\odot}
{\rm yr}^{-1},
\label{gasmass}
\end{equation}
where $n_{\rm He/H}$ is the total number density of Helium/Hydrogen,
$T_{\rm mb}$ is the main-beam brightness temperature, $v_{e}$ represent
the gas expansion velocities, $D$ is the distance to the source,
$A(B,J)$ is a quantity that depends on the beam size used, $B$, and the
transition observed, $J \rightarrow J-1$, and $f_{\rm CO}$ is the
abundance of CO with respect to H$_{2}$.
Sch\"oier \& Olofsson (\cite{schoier+olofsson01}) 
find that the gas mass-loss rates resulting from  
Eq.\,(\ref{gasmass}) are systematicly underestimated compared to a more
detailed radiative transfer analysis (see their Fig.\,8), however, for
these particular stars the values agree resonable well.

In the models discussed here we obtain higher gas-to-dust ratios
than the values given by Olofsson et al.\  (\cite{olofsson+etal93}).
This could be a consequence of the fact that some of the stellar
parameters differ from these of the models. As mentioned above,
the models and stars for the comparison were chosen for similar
wind velocities and dust mass-loss rates, not for comparable
stellar parameters. In this sense, we are rather comparing the
observable properties of the circumstellar dust shells than of
the stars and their surrounding gas envelopes. Therefore,
the comparison shown here should not be considered as an attempt
to fit observations of individual stars but to investigate
certain wind properties of our models.

While the average colors of the models are comparable to the average
colors of the observed stars, see Fig.\,\ref{massloss_color}, the temporal
variation of the colors for a given model is smaller than the variations
observed in the individual stars. This is probably a consequence of
relatively small bolometric luminosity variations in the models.

Figure \ref{massloss_color} shows a comparison of models and observed stars 
for the colors J-H vs.\ H-K. A qualitatively similar relation between 
models and observations is obtained for the colors K-L vs.\ H-K.


\section{Conclusions}

We have investigated in detail how the predicted wind properties 
of carbon-rich AGB stars are influenced by the choice of 
micro-physical dust parameters i.e.\ the optical properties of 
the dust, the intrinsic dust density, the assumed sticking 
coefficients and the surface tension of the grain material (these two 
last parameters control the efficiency of the dust formation).

For the theoretical predictions of mass-loss it is important to know
how the uncertainty in the chosen dust parameters 
affects the obtained results.
Varying the micro-physical parameters within the range typical of possible
materials can change the 
value for the mean outflow velocity of the gas and dust  
as well as the predicted degree of dust condensation by a factor of ten and the 
predicted mass-loss by a factor of four. In the transition region between models 
with and without mass-loss the choice of micro-physical parameters  
is vital for whether a particular set of stellar
parameters will give rise to a dust-driven mass-loss or not. 

The main source of momentum for the stellar wind is the radiation pressure on
dust. The radiation pressure on the dust and the radiative equilibrium grain
temperature is determined by the wavelength dependence of the grain extinction
efficiency. The steeper the dependence of wavelength, the larger the 
difference between the equilibrium grain temperature and the radiation
temperature. The radiation pressure on the other hand is proportional to the
flux mean opacity which both depends on the slope of the extinction
efficiency as a function of wavelength and on its absolute value. The 
latter may differ by almost an order of magnitude for different types
of amorphous carbon in the critical region around 1~$\mu$m. The density
of the grain material has to be chosen consistently with the grain extinction
efficiency.

The surface tension of the grain material and the sticking coefficients are very significant for the
calculated rates at which grains are formed out of the gas (nucleation)
and at which new material is added to existing grains (grain growth). Even
a moderate variation of the values within the range expected for possible 
materials has noticeable consequences for the properties of the dust-driven
stellar winds, including the resulting near-infrared colors.

The colors are similar to stars with comparable dust mass-loss rates and outflow
velocities, however, we have not attempted to fit any individual stars.


\begin{acknowledgements}
      The authors would like to thank Kjell Eriksson for valuable
      discussions.
      ACA gratefully acknowledges support from the Carlsberg Foundation.
      This work was supported by NorFA, the Swedish Research 
      Council (VR) and the Royal Swedish Academy of Sciences (KVA). 
\end{acknowledgements}



\begin{thebibliography}{}

   \bibitem[]{} 

     \bibitem[1973]{allen73} Allen C.W., 1973, Astrophysical Quantities, The Athlone
       Press, London

    \bibitem[1999]{andersen+etal99} Andersen A., Loidl R. \& H\"ofner S., 1999, A\&A  349, 243

   \bibitem[1935]{becker+doring35} Becker R. \& D\"oring W., Ann.\ Physik, A125, 719

   \bibitem[1983]{bohren+huffman83} Bohren C.F. \& Huffman D.R., 
       1983, Absorption and Scattering
       of Light by Small Particles (John Wiley \& Sons, New York)

   \bibitem[1987]{bussoletti+etal87}Bussoletti E., Colangeli L., Borghesi A. \& Orofino V., 1987,
       A\&AS 70, 257

   \bibitem[1976]{campbell+etal76} Campbell M.F., Harvey P.M., Hoffmann W.F., et al., 
1976, ApJ 208, 396

    \bibitem[1985]{donn+nuth85} Donn B. \& Nuth J.A.III., 1985, ApJ 288 187

   \bibitem[1995]{dorfi+feuchtinger95} Dorfi E. \& Feuchtinger M., 1995, Computer Physics Communications 89, 69 

    \bibitem[1991]{dorfi+hoefner91} Dorfi E.A. \& H\"ofner S., 1991, A\&A 248, 105

    \bibitem[1984]{draine+lee84} Draine B.T. \& Lee H.M., 1984, ApJ 285, 89

     \bibitem[1980]{dreiling+bell80} Dreiling L.A. \& Bell R.A., 1980, ApJ 241, 736

    \bibitem[1966]{feder+etal66} Feder J., Russel K.C., Lothe J. \& Pound G.M., 1966, Advances in
      Physics 15, 111

    \bibitem[1992]{fleischer+etal92} Fleischer A.J., Gauger A. \& Sedlmayr E., 1992, A\&A 266, 321

   \bibitem[1984]{gail+sedlmayr84} Gail H.-P. \& Sedlmayr E., 1984, A\&A 132, 163

   \bibitem[1986]{gail+sedlmayr86} Gail H.-P. \& Sedlmayr E., 1986, A\&A 166, 225

   \bibitem[1987a]{gail+sedlmayr87a} Gail H.-P. \& Sedlmayr E., 1987a, A\&A 171, 197

   \bibitem[1987b]{gail+sedlmayr87b} Gail H.-P. \& Sedlmayr E., 1987b, A\&A 177, 186

   \bibitem[1988]{gail+sedlmayr88} Gail H.-P. \& Sedlmayr E., 1988, A\&A 206, 153

    \bibitem[1990]{gauger+etal90} Gauger A., Gail H.-P. \& Sedlmayr E., 1990, A\&A 235, 345

    \bibitem[1995]{gurtler+etal96} G\"urtler J., K\"ompe C. \& Henning Th., 1996, A\&A 305, 878

   \bibitem[1986]{hale86} Hale B.N., 1986, Phys.\ Rev.\ A 33, 4156

    \bibitem[1989]{hale+etal89} Hale B.N., Kemper P. \& Nuth J.A.III., 1989, J.\ Chem.\ Phys.\ 91, 4314

    \bibitem[1999]{hoefner99} H\"ofner S., 1999, A\&A 346, L9

    \bibitem[1992]{hoefner+dorfi92} H\"ofner S. \& Dorfi E.A., 1992, A\&A 265, 207

    \bibitem[1997]{hoefner+dorfi97} H\"ofner S. \& Dorfi E.A., 1997, A\&A 319, 648

   \bibitem[1995]{hoefner+etal95} H\"ofner S., Feuchtinger M.U. \& Dorfi E.A., 1995, A\&A 279, 815

   \bibitem[2002]{hoefner+etal02} H\"ofner S., Gautschy-Loidl R., Aringer B. \& J{\o}rgensen U.G., 2002, A\&A in press

   \bibitem[1988]{huffmann88} Huffmann D.R., 1988, in Experiments on Cosmic Dust Analogues, Kluwer Academic Publishers, eds. E.\,Bussoletti, C.\,Fusco, G.\,Longo, 25 
   \bibitem[1998]{jaeger+etal98} J\"ager C., Mutschke H. \& Henning Th., 1998, A\&A 332, 291

    \bibitem[1987]{jura87} Jura M., 1987, ApJ 313, 743

    \bibitem[1997]{joergensen97} J{\o}rgensen U.G., 1997, In: van Dishoeck E.F. (ed.) Mulecules
    in Astrophysics: Probes and processes, [Kluwer], IAU Symp.\ 178, 441

    \bibitem[1985]{knapp+morris85} Knapp G.R. \& Morris M., 1985, ApJ 292, 640

    \bibitem[1968]{landolt-bornstein68} Landolt-B\"ornstein, 1968,
      Zahlenwerte und Funktionen, vol. 5b, Springer Verlag, Berlin

   \bibitem[1987]{martin+rogers87} Martin P.G. \& Rogers C., 1987, ApJ 322, 374

   \bibitem[2001]{martinez+etal01} Martinez D.M., Ferguson F.T., Heist R.H. \& Nuth J.A.III., 2001, J.\ Chem.\ Phys.\ 115, 310

    \bibitem[1990]{maron90} Maron N., 1990, Ap\&SS 172, 21

    \bibitem[1993]{nuth+ferguson93} Nuth J.A. \& Ferguson F., 1993, Ceramic Transactions 30, 23

    \bibitem[1993]{olofsson+etal93} Olofsson H., Eriksson K., Gustafsson B. \& Carlstr\"om U., 1993, ApJS 87, 267

     \bibitem[1972]{pound72} Pound G.M., 1972, J.\ Phys.\ Chem.\ Ref.\ Data 1, 135

     \bibitem[1991]{rouleau+martin91} Rouleau F. \& Martin P.G., 1991, ApJ 377, 526

   \bibitem[1973]{salpeter73} Salpeter E.E., 1973, J.\ Chem.\ Phys.\ 58, 4331

 \bibitem[2001]{schoier+olofsson01} Sch\"oier F.L. \& Olofsson H., 2001, A\&A 368, 969 

   \bibitem[1994]{sedlmayr94} Sedlmayr E., 1994, in Molecules in the Stellar Environment,
     LNP 428, ed.\ U.G. J{\o}rgensen (Springer, Berlin), 163

   \bibitem[1985]{sopka+etal85} Sopka R.J., Hildebrand R.H., Jaffe D.T., et al.,
1985, ApJ 294, 242

  \bibitem[1971]{stull+prophet71} Stull D.R. \& Prophet H., 1971, JANAF Thermochemical 
      Tables, 2nd Ed., Nat.\ Bureau of Standards, Washington, (NSRDS-NBS37)

   \bibitem[1975]{tabak+etal75} Tabak R.G., Hirth J.P., Meyrick G. \& Roark T.P., 1975, ApJ 196, 457

   \bibitem[1997]{whitelock+etal97} Whitelock P., Feast M., Marang F. \& Overbeek M., 1997, MNRAS 288, 512  

\end{thebibliography}
\end{document}